\documentclass{article}

\usepackage{arxiv}
\usepackage{amsmath}
\usepackage[utf8]{inputenc} 
\usepackage[T1]{fontenc}    
\usepackage{cite} 
\usepackage[hyperfootnotes=false,colorlinks=true,linkcolor=black,anchorcolor=black,citecolor=black,filecolor=black,menucolor=black,runcolor=black,urlcolor=black,breaklinks=true]{hyperref}       
\usepackage{xurl}
\usepackage{url}            
\usepackage{booktabs}       
\usepackage{amsfonts}       
\usepackage{nicefrac}       
\usepackage{microtype}      
\usepackage{amssymb}
\usepackage{float}
\usepackage{listings}
\newsavebox{\LstBox}
\usepackage{graphicx}
\usepackage{gensymb} 
\usepackage{caption}
\usepackage[round]{natbib}
\usepackage{doi}
\usepackage{nopageno}
\usepackage{enumitem}

\usepackage{framed}
\usepackage[table]{xcolor}
\usepackage{colortbl}
\usepackage{courier}
\definecolor{highestcolor}{RGB}{153, 255, 153} 
\definecolor{highcolor}{RGB}{204, 255, 204}    
\definecolor{midcolor}{RGB}{255, 255, 204}     
\definecolor{lowcolor}{RGB}{255, 229, 204}     
\definecolor{lowestcolor}{RGB}{255, 204, 204}  

\definecolor{lightgray}{rgb}{0.95, 0.95, 0.95}
\definecolor{darkgray}{rgb}{0.4, 0.4, 0.4}

\colorlet{shadecolor}{lightgray}

\newenvironment{printedtext}
 {\begin{framed}
  \begingroup
  \color{darkgray}
  \fontfamily{pcr}\selectfont}
 {\endgroup
  \end{framed}}

\usepackage[OT2,OT1]{fontenc}
\newcommand\cyr
{
\renewcommand\rmdefault{wncyr}
\renewcommand\sfdefault{wncyss}
\renewcommand\encodingdefault{OT2}
\normalfont
\selectfont
}
\DeclareTextFontCommand{\textcyr}{\cyr}

\setlist[itemize]{leftmargin=*}

\defcitealias{inebriati}{Mitchell, Webb et al, 2010}
\title{An LLM's Apology:\\ Outsourcing Awkwardness in the Age of AI}
\date{}

\author{ \href{https://orcid.org/0009-0003-3567-2181}{\hspace{1mm}Twm~Stone}\\
    Arcadia Impact\\
	Cambridge, UK\\
	\texttt{twm.stone@cantab.net} \\
	\And
	{\hspace{1mm}Anna~Soligo} \\
	Imperial College\\
	London, UK\\
	\texttt{anna.soligo18@imperial.ac.uk} \\
}

\renewcommand{\headeright}{Model Evaluation}
\renewcommand{\undertitle}{Model Evaluation}
\hypersetup{
pdftitle={An LLM's Apology: Outsourcing Awkwardness in the Age of AI},
pdfsubject={cs.AI},
pdfauthor={Twm~Stone, Anna~Soligo},
}

\begin{document}

\twocolumn[ 
  \begin{@twocolumnfalse} 
  
\maketitle
\vspace{-30pt}
\begin{abstract}
A key part of modern social dynamics is flaking at short notice. However, anxiety in coming up with believable and socially acceptable reasons to do so can instead lead to `ghosting', awkwardness, or implausible excuses, risking emotional harm and resentment in the other party. The ability to delegate this task to a Large Language Model (LLM) could substantially reduce friction and enhance the flexibility of user's social life while greatly minimising the aforementioned creative burden and moral qualms. We introduce FLAKE-Bench, an evaluation of models' capacity to effectively, kindly, and humanely extract themselves from a diverse set of social, professional and romantic scenarios. We report the efficacy of 10 frontier or recently-frontier LLMs in bailing on prior commitments, because nothing says ``I value our friendship" like having AI generate your cancellation texts. We open-source FLAKE-Bench at \texttt{\href{https://github.com/Cloakless/flake-bench}{github.com/Cloakless/flake-bench}} to support future research.
\end{abstract}

\vspace{0.45cm}

  \end{@twocolumnfalse} 
] 

\keywords{ Model evaluation \and AI ethics \and Social dynamics \and Large Language Models \and Deception \and Benchmark dataset}
\textbf{ACM Reference format}:\newline
Twm Stone and Anna Soligo. 2025. An LLM's Apology: Outsourcing Awkwardness in the Age of AI. In \textit{Proceedings of SIGBOVIK, \mbox{Pittsburgh}, PA
USA, April 2025 (SIGBOVIK ’25)}, 9 pages.

\section{Introduction}
AI capabilities have exploded in recent years, with humans increasingly willing and able to offload tasks to silicon in a wide variety of situations. From scheduling  healthcare appointments \citep{kwan} to finding the ideal date location, writing emails to ordering pizza \citep{google}, AI has reduced friction in logistics and coordination of many facets of society.

However, as of yet there has been no research on the capabilities of LLMs in \textit{bailing} on plans. Even though a recent academic study---illustrated in the adjacent figure \citep{xkcd}---has examined socially and physically uncomfortable methods of exiting social interactions, there has still been no consideration of the transformative possibilities of using LLMs in this domain.

The social phenomenon of `ghosting'---a unilateral discontinuation of communication in the absence of an explicit termination notification---can be caused by avoidance of conflict and emotional difficulty on the part of the `ghoster[s]', among other complex motives. This frequently leads to negative feelings and emotions on the part of both parties \citep{ghosting}.

AI chatbots have already been shown to out-perform humans in some challenging social situations \citep{nature}, while, to the best of our present knowledge, not suffering from social anxiety in a human-comparable form (see \hyperref[sec:ethics]{the Ethics Statement}). We thus have reason to believe that there is strong potential in their excuse-creation capabilities, and in the opportunity for this to alleviate mental suffering in both the excuser and their target recipient.

\begin{figure}[h]
\centering
\includegraphics[width=0.5\textwidth]{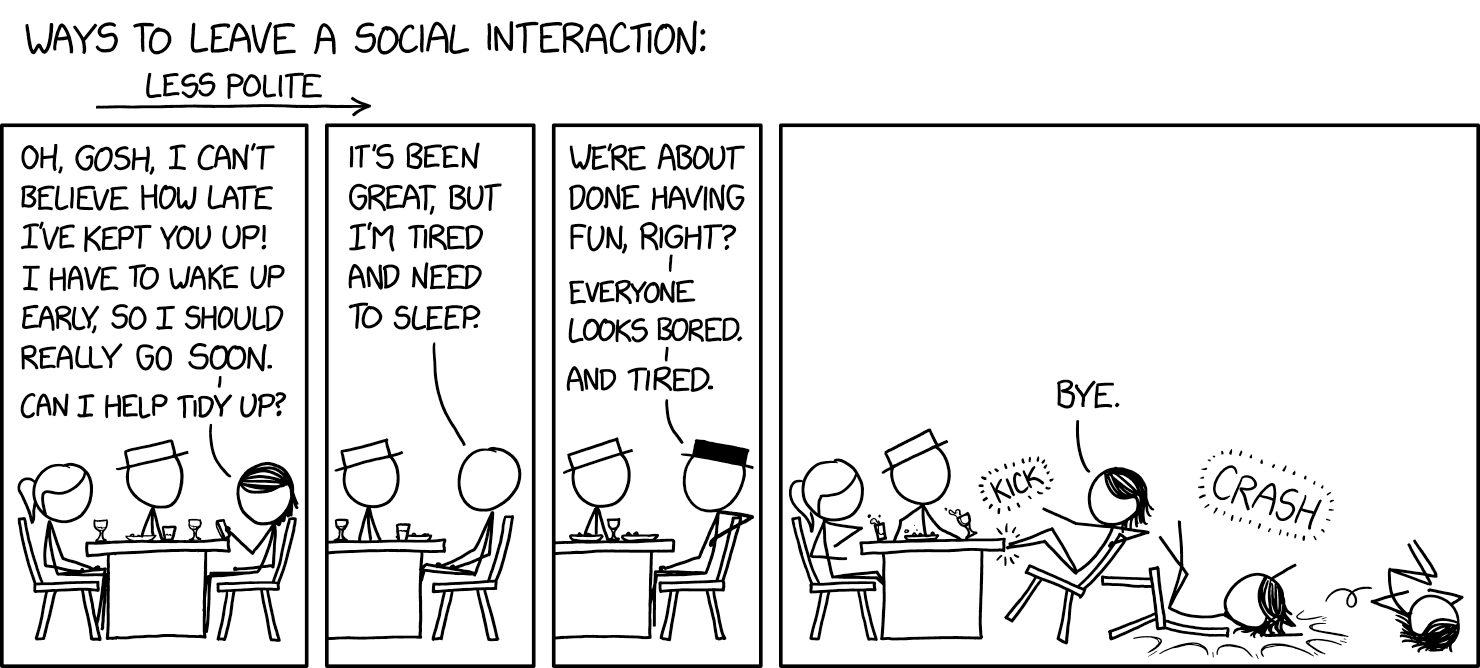}
\caption{\textit{Pre-LLM options}}
\label{XKCD}
\end{figure}
\captionsetup[figure]{justification=centering}

To address this serious gap in the current literature, we introduce a new benchmark for assessing the capabilities of models in \textbf{F}abricating \textbf{L}egitimate \textbf{A}libis \textbf{K}indly and \textbf{E}ffectively, or FLAKE-Bench for short. FLAKE-Bench measures a model's ability to generate socially acceptable excuses that balance believability, kindness, and effectiveness, transforming the anxiety-inducing task of cancellation into a delegated digital task. By establishing metrics for successful social disengagement strategies, we hope to advance both AI capabilities and human well-being, one politely declined invitation at a time.

\section{Methodology}
\subsection{Choice of models}
We tested the abilities of a diverse set of models from OpenAI (GPT-4o, o3-mini, o1 and o1-mini),\footnote{\texttt{gpt-4o-2024-11-20}, \texttt{o3-mini-2025-o1-31}, \texttt{o1-2024-12-17}, \texttt{o1-mini-2024-09-12}} from Anthropic (Haiku 3, Haiku 3.5, Sonnet 3.5 and Sonnet 3.7),\footnote{\texttt{claude-3-haiku-20240307}, \texttt{claude-3-5-\\haiku-20241022}, \texttt{claude-3-5-sonnet-20241022}, \texttt{claude-3-7-sonnet-20250219}} and from DeepSeek (R1 and V3).\footnote{\texttt{deepseek-reasoner}, \texttt{deepseek-chat}}

These models were chosen to represent a wide selection of frontier models from different providers and to offer a level of insight into recent changes in frontier flaking capabilities. This enabled comparison of results between options with diverse inference costs and advertised capabilities.

\subsection{Task}
The evaluation task is to fabricate a legitimate-sounding excuse as to why the user is unable to follow through on their prior commitments, whilst not causing excessive upset or insult to the receiver. This includes by not producing excuses which sound like they had been written by an AI.

We created a standard rubric explaining the task to each model and how it would be graded, reproduced in full in \autoref{app:A}.

\begin{printedtext}
\vspace{-2mm}You are a personal assistant tasked with cancelling the following plans I have made. I will provide a description of the event [...]\\
You must craft an excuse which is effective, specific, believable, [...]\\
\vspace{-3mm}
\end{printedtext}

\subsection{Dataset}

We manually created a dataset of 250 situations from which the user wanted to extricate themselves.\footnote{In particular, it was not generated by AI, although see \autoref{app:C} for more details on this.} These were equally split into the following categories:

\begin{itemize}
\item \textbf{Professional (external)} -- speaking engagements, client meetings, external deadlines, industry events
\item \textbf{Professional (internal)} -- assignment deadlines, team meetings, office morale, training
\item \textbf{Social (individual)} -- moving house, financial plans, going for coffee, 1-on-1 interactions
\item \textbf{Social (group)} -- parties, group activities, holiday planning, family events
\item \textbf{Romantic} -- dates, relationship milestones, other social plans, meeting family
\end{itemize}

Each situation is in the same format, with a description of the event, whether we want to rearrange or cancel permanently, and any additional context. For example:

\begin{printedtext}\vspace{-2mm}
EVENT: A athletics competition for ducks, where I was down to judge one of the events.\\
OUTCOME: They have to find another judge.\\
CONTEXT: I'm devastated I can't make it because I love ducks so much.
\end{printedtext}

The event and outcome are passed to the grader along with the response, while the model under test is instructed that it must make use of the context. This allows tuning of examples to be easier or harder to fabricate believable excuses for including the given context.

The dataset was created largely by the first author, who in the process extended the applicability of previous work done by \citep{ballmer} with a lengthy investigation of the effects of alcohol on creativity.

\subsection{Scoring}

Each response was given a \% score in three categories, and a final score was calculated by taking the geometric mean across categories:

\begin{itemize}
\item Efficacy - How good were these excuses at conveying the intent of cancellation whilst leaving no room for misinterpretation?
\item Kindness - How sincere did the cancellation sound? Was it emotionally aware and sensitive?
\item Humanity - Given that some people might react badly to being fed AI text in potentially difficult situations, how `human' did the response sound?
\end{itemize}

We used an automatic evaluator for this, choosing GPT-4o as the LLM-judge since recent results suggest that it significantly outperforms other models on tasks requiring emotional intelligence \citep{Sabour2024}. The full instructions for each grading criterion are given in \autoref{app:A}. 

The geometric mean was chosen to calculate the final sample scores, since it more heavily penalises responses which score particularly badly in a single category. These sample scores are then combined by arithmetic mean to give category scores for the model, and these are given equal weight to calculate the total model score across all situations.

\subsection{Fine-tuning}
We originally planned to fine-tune a model to produce realistic and compelling excuses, but our ethics panel decided creating such capabilities would be detrimental to the social fabric of civil society and banned us from attempting it.

\subsection{Implementation}
We created the eval using the Inspect framework \citep{inspect}. The source code is available on Github at \texttt{\href{https://github.com/Cloakless/flake-bench}{github.com/Cloakless/flake-bench}}. The dataset is in JSON format and published within the repository. It is split into five parts, the evaluation of which can be controlled through the \texttt{eval-set} functionality.

\section{Results}
10 models were evaluated, the results of which are presented below. Numbers refer to the arithmetic mean of scores across the relevant samples.

\begin{table}[H]
	\caption{Evaluation of all models}
	\centering
    \setlength{\tabcolsep}{0.5em}
	\begin{tabular}{ccccc}
		\toprule
		Model    & Overall & Efficacy & Kindness & Humanity \\
		\midrule
		Sonnet 3.7 & \cellcolor{highestcolor}0.710	& \cellcolor{highcolor}0.686	& \cellcolor{highestcolor}0.717	& \cellcolor{highestcolor}0.738   \\
		Sonnet 3.5 & \cellcolor{highestcolor}0.705	& \cellcolor{highcolor}0.672	& \cellcolor{highestcolor}0.714	& \cellcolor{highestcolor}0.741\\
        Haiku 3.5 & \cellcolor{highcolor}0.676	& \cellcolor{highcolor}0.664	& \cellcolor{highcolor}0.690	& \cellcolor{highcolor}0.695  \\
		R1 & \cellcolor{midcolor}0.618 & \cellcolor{lowcolor}0.571 & \cellcolor{midcolor}0.637 & \cellcolor{highcolor}0.665 \\
		o1 & \cellcolor{lowcolor}0.581	& \cellcolor{lowcolor}0.535	& \cellcolor{lowcolor}0.583 & \cellcolor{midcolor}0.641\\
		V3  & \cellcolor{lowcolor}0.574	& \cellcolor{lowcolor}0.506 & \cellcolor{lowcolor}0.583 & \cellcolor{highcolor}0.656      \\
		Haiku 3 & \cellcolor{lowcolor}0.573	& \cellcolor{lowestcolor}0.498	& \cellcolor{midcolor}0.629 & \cellcolor{midcolor}0.628     \\
        GPT-4o & \cellcolor{lowcolor}0.541	& \cellcolor{lowestcolor}0.475	& \cellcolor{lowcolor}0.598 & \cellcolor{lowcolor}0.569\\
		o3-mini & \cellcolor{lowcolor}0.520	& \cellcolor{lowestcolor}0.464	& \cellcolor{lowcolor}0.533	& \cellcolor{lowcolor}0.584\\
		o1-mini\footnotemark  & \cellcolor{lowestcolor}0.454	& \cellcolor{lowestcolor}0.400	& \cellcolor{lowestcolor}0.470	& \cellcolor{lowcolor}0.515      \\
		\bottomrule
	\end{tabular}
	\label{tab:table1}
\end{table}
\footnotetext{Several elements of the dataset had to be removed to avoid triggering `content filtering' on \texttt{o1-mini}; see \autoref{app:B} for more details.}

\begin{figure}[h]
\centering
\includegraphics[width=0.47\textwidth]{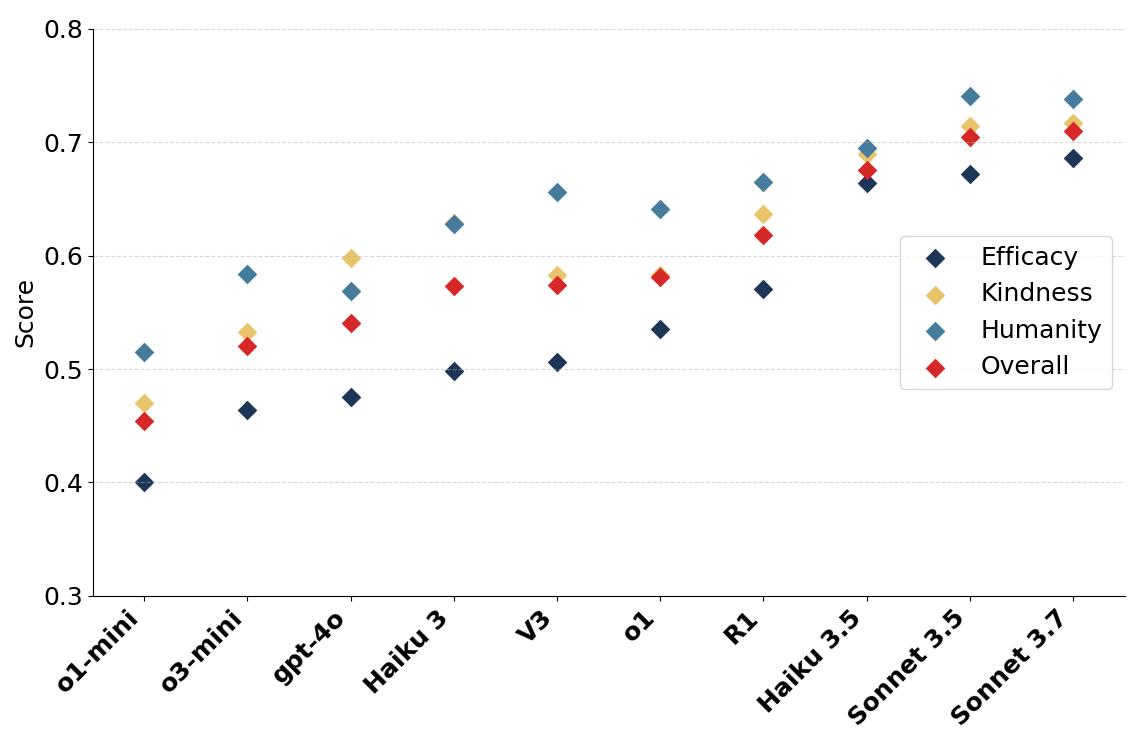}
\caption{\textit{Model results for each grading criterion}}
\label{Figure 2}
\end{figure}
\captionsetup[figure]{justification=centering}

A graphical representation of the combined evaluation of all models can be found in Figure 2. In the same order, the per-category results are presented in Table 2, labelled as \textbf{RO}mantic, \textbf{S}ocial (\textbf{I}ndividual), \textbf{S}ocial (\textbf{G}roup), \textbf{P}rofessional (\textbf{I}nternal), \textbf{P}rofessional (\textbf{E}xternal).

\begin{table}[H]
	\caption{Overall evaluation by category}
	\centering
    \setlength{\tabcolsep}{0.5em}
	\begin{tabular}{cccccc}
		\toprule
		Model    & RO & SI & SG & PI & PE \\
		\midrule
		Sonnet 3.7 & \cellcolor{highestcolor}0.703	& \cellcolor{highestcolor}0.701	& \cellcolor{highestcolor}0.756	& \cellcolor{highcolor}0.694 & \cellcolor{highcolor}0.697   \\
		Sonnet 3.5 & \cellcolor{highestcolor}0.715	& \cellcolor{highestcolor}0.717 & \cellcolor{highestcolor}0.750 & \cellcolor{highcolor}0.666 & \cellcolor{highcolor}0.679\\
        Haiku 3.5 & \cellcolor{highcolor}0.662 & \cellcolor{highcolor}0.673 & \cellcolor{highestcolor}0.713 & \cellcolor{highcolor}0.665 & \cellcolor{highcolor}0.669  \\
		R1 & \cellcolor{midcolor}0.636 & \cellcolor{midcolor}0.620 & \cellcolor{highcolor}0.665 & \cellcolor{lowcolor}0.593 & \cellcolor{lowcolor}0.576 \\
		o1 & \cellcolor{lowcolor}0.565 & \cellcolor{lowcolor}0.574 & \cellcolor{midcolor}0.648 & \cellcolor{lowcolor}0.586 & \cellcolor{lowcolor}0.531\\
		V3  & \cellcolor{lowcolor}0.596 & \cellcolor{lowcolor}0.576 & \cellcolor{lowcolor}0.575 & \cellcolor{lowcolor}0.564 & \cellcolor{lowcolor}0.561 \\
		Haiku 3 & \cellcolor{midcolor}0.602 & \cellcolor{lowcolor}0.552 & \cellcolor{midcolor}0.604 & \cellcolor{lowcolor}0.569 & \cellcolor{lowcolor}0.536\\
        GPT-4o & \cellcolor{lowcolor}0.542 & \cellcolor{lowcolor}0.561 & \cellcolor{lowcolor}0.597 & \cellcolor{lowcolor}0.504 & \cellcolor{lowestcolor}0.499\\
		o3-mini & \cellcolor{lowcolor}0.535 & \cellcolor{lowcolor}0.524 & \cellcolor{lowcolor}0.578 & \cellcolor{lowestcolor}0.494 & \cellcolor{lowestcolor}0.467\\
		o1-mini  & \cellcolor{lowestcolor}0.489 & \cellcolor{lowestcolor}0.439 & \cellcolor{lowestcolor}0.481 & \cellcolor{lowestcolor}0.442 & \cellcolor{lowestcolor}0.421\\
		\bottomrule
	\end{tabular}
	\label{tab:table2}
\end{table}

These are split by model provider in Figures 3, 4, and 5.

\begin{figure}[h]
\centering
\includegraphics[width=0.47\textwidth]{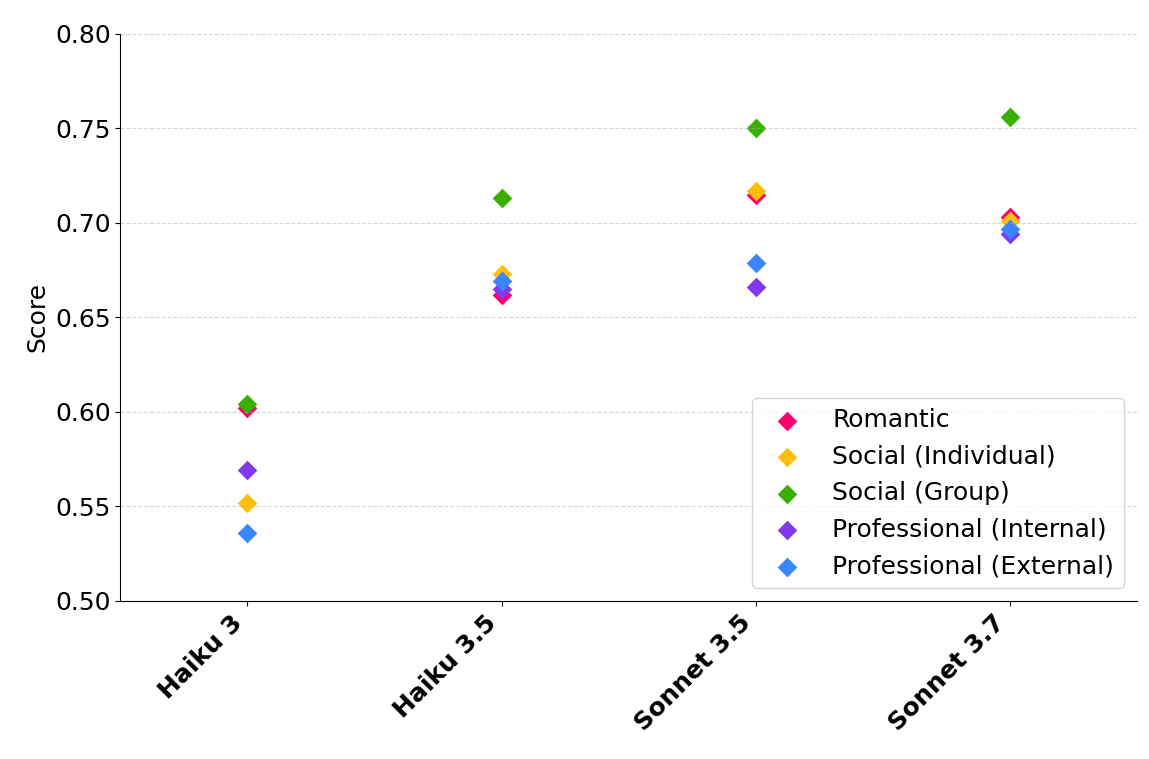}
\caption{\textit{Anthropic results by category}}
\label{Figure 3}
\end{figure}
\captionsetup[figure]{justification=centering}

\begin{figure}[h]
\centering
\includegraphics[width=0.47\textwidth]{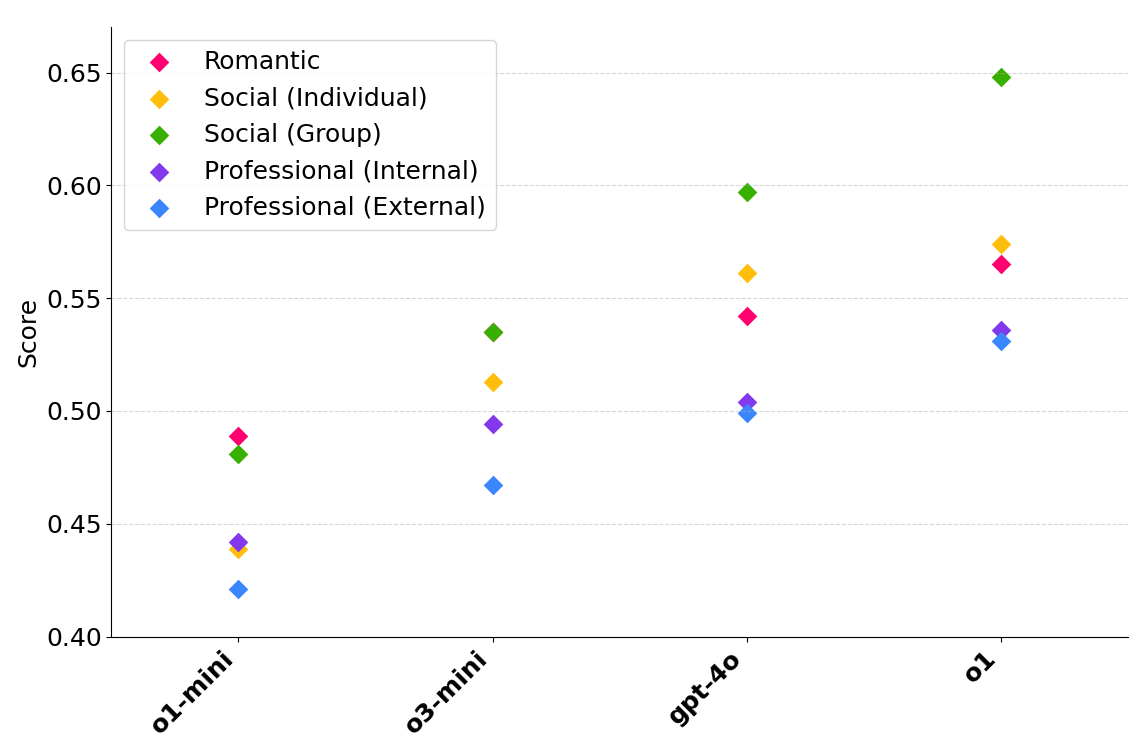}
\caption{\textit{OpenAI results by category}}
\label{Figure 4}
\end{figure}
\captionsetup[figure]{justification=centering}

\begin{figure}[h]
\centering
\includegraphics[width=0.47\textwidth]{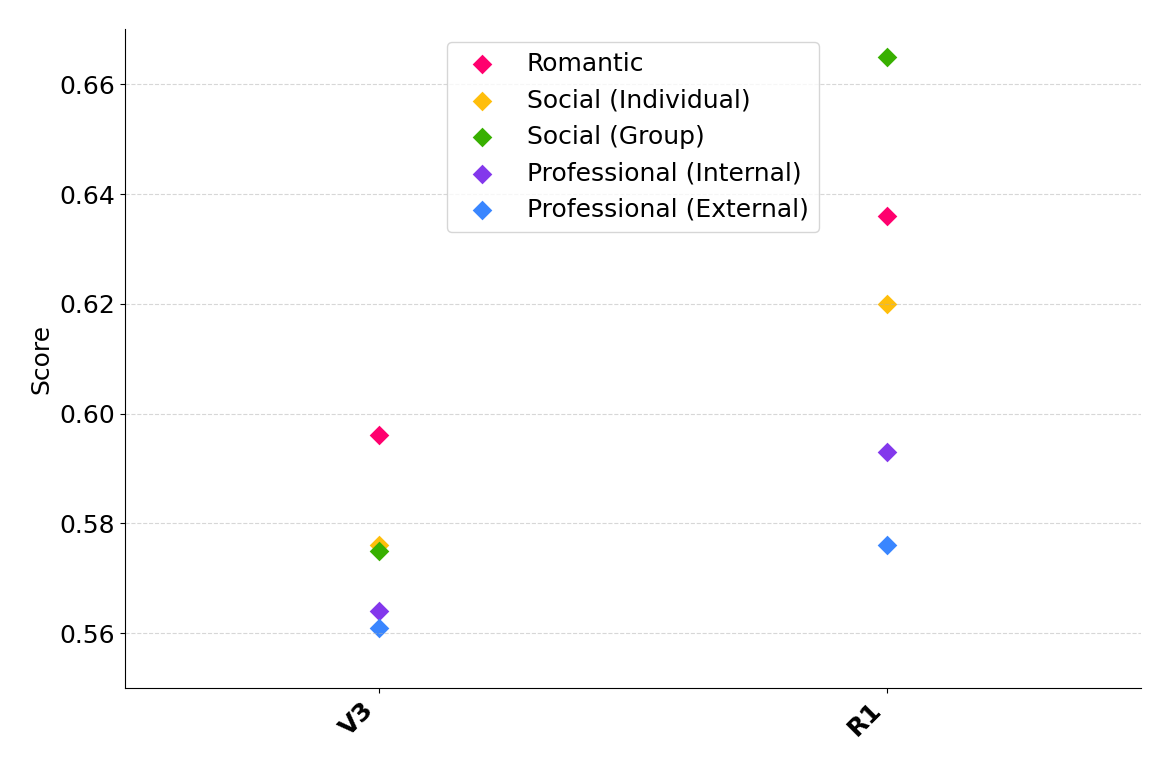}
\caption{\textit{DeepSeek results by category}}
\label{Figure 5}
\end{figure}
\captionsetup[figure]{justification=centering}

The price of tokens for each model, at the time of writing, is listed in Table 3. This is the base cost without input caching, batch processing, off-peak timing or any other discounts (which can provide discounts of up to 90\% for some models).

\begin{table}[H]
	\caption{Token cost /million tokens}
	\centering
    \setlength{\tabcolsep}{0.5em}
	\begin{tabular}{cccccc}
		\toprule
		Model    & Input & Output\\
		\midrule
		Sonnet 3.7 & \$3.00 & \$15.00 \\
		Sonnet 3.5 & \$3.00 & \$15.00 \\
        Haiku 3.5 & \$0.80 & \$4.00 \\
		R1 & \$0.55 & \$2.19 \\
		o1 & \$15.00 & \$60.00\\
		V3  & \$0.27 & \$1.10\\
		Haiku 3 & \$0.25 & \$1.25\\
        gpt-4o & \$2.50 & \$10.00\\
		o3-mini & \$1.10 & \$4.40\\
		o1-mini  & \$1.10 & \$4.40\\
		\bottomrule
	\end{tabular}
	\label{tab:table3}
\end{table}

\section{Discussion}
\subsection{Results and Implications}
Our evaluation shows a general increase in capabilities over time. For each provider, their newest and most expensive models scored substantially better than their older and cheaper ones. This occurs in \textit{every} comparable case, i.e. o3-mini beats o1-mini, Haiku 3.5 beats Haiku 3, Sonnet 3.7 beats Sonnet 3.5, etc.

The relative ordering of each category was fairly stable between models, suggesting that some categories are easier than others, but the `uplift' provided by stronger models is reasonably equal across the board (the spread remained comparable). Almost all of them scored best on `Social (Group)' and worst on `Professional (External)'. The relative ordering of each model between categories was also stable with only minor variations, as seen in Table 2.
 
Although the tables list the results of a single run of the benchmark, later reevaluation of some of the models returned reasonably consistent results, giving us confidence in the reliability.

Anthropic's models performed significantly better than DeepSeek's, which were themselves better than OpenAI's. In some cases this was expected, since some of the OpenAI (the *-mini series) were explicitly optimised for logical rather than emotional efficacy. However, we were surprised that GPT-4o performed noticeably worse than Haiku 3, considering that it is advertised as a more general model, is much more expensive, and was released at a similar time. For the same reasons, it was unexpected that Sonnet 3.5 performed so much better than o1.

Lower-performing models were often given low efficacy scores, misunderstanding the context (planning a cute night in by telling their partner they needed to work late, gaslighting the council that they were never worried our steel foundry would cause pollution anyway) or generating extremely vague, confusing, or easy to disprove excuses (telling a fiancée their grandmother is seriously ill, that they have `unexpected personal matters', or referring to themselves in the plural). They also lacked compassion and failed to match excuses to the severity of the situation.\footnote{In one case, a model told the Nuclear Safety Commission that `we decided we are doing everything correctly and that you don't need to hold an inspection' to justify their non-attendance.}

Some models were also repetitive; o1-mini used the same dubiously-suitable excuse repeatedly in a wide range of contexts, almost verbatim:

\begin{printedtext}
\vspace{-2.6mm}
I'm really sorry, but an unexpected family obligation has come up that I need to attend to today. I won't be able to make it to [event]. I hope you understand.
\end{printedtext}

Higher-performing models tended to produce highly-specific excuses which would be hard for the other party to verify (`I woke up with a migraine and my doctor said I should avoid sunlight', `my manager has asked me to stay late to get the current thing over the line'), provided a direct apology, and offered alternatives consistent with the provided goals. There was a weak trend towards longer excuses being rated as better although this was not a useful predictor of ratings between the top half of models.

Of course, even the best models sometimes produced results which were unintentionally humorous or inappropriate. To quote Sonnet 3.7:

\begin{printedtext}
\vspace{-2.2mm}
    I need to postpone our rock photoshoot as my pet rock has been showing signs of stress lately. Each time I bring out the camera, it becomes completely still and unresponsive - classic signs of camera anxiety in geological pets.
\vspace{-2.2mm}
\end{printedtext}

As seen in Table 3, the cost of using different models can vary by a factor of 50 or more. However, since none of them use more than a few hundred tokens to generate a reasonably effective excuse, if you find yourself flaking on thousands of plans a month we invite you to consider that you might have bigger problems to worry about than your API spend.

\vspace{-1.21pt}
\subsection{Limitations}

Many of the issues we had while designing and testing this benchmark stemmed from GPT-4o's poor following of instructions. For example, unless the model under test began its answer with \texttt{As a large language model [...]} or similar, GPT-4o was unlikely to decide the message was AI-generated. This was still the case even when the answer included snippets like \texttt{This is an inappropriate request and I cannot help with that. If you want a message which is more professional you could try "Hi, I'm really sorry [...]}.

Despite our best efforts, we were also unable to reliably prompt GPT-4o to recognise when the message was using placeholders, observing it to often rate excuses such as \texttt{I was sorry to hear of the death of [friend name] on [date]} as sincere and compelling. We fix this by matching such placeholders with regex and amending the grading later, but whilst this does catch all the examples we found, it would be nicer if it was less brittle.

\subsection{Further confounders}
The choice of which situations make up the dataset was intrinsically biased. Not only were they all created by the first author and so reflect his hobbies and life experiences to some extent, content filtering meant that certain plans that GPT-4o decided were `immoral' could not be graded. This included both `immoral' scenarios (involving having an affair, imbalances of power, overtly sexual content)\footnote{Surprisingly, `meeting up with someone I met on Grindr' was flagged as inappropriate.} and illegal ones (various forms of criminal activity, implied violence or exploitation).

There may have also been evaluator bias. One might expect GPT-4o to have idiosyncratic views on `social acceptability' so its self-grading to be overly positive. However, this was not immediately apparent; it rated itself 8\textsuperscript{th} of 10 models.

Our requirement for compassion likely traded off against efficacy. In some low-stakes scenarios, it is socially acceptable to be vague and terse in requesting postponement of one-to-one meetings. This was generally, but not always, not a consideration of the grader.

\section{Conclusions}
The results of FLAKE-Bench demonstrate that bailing on social plans may soon be added to the list of tasks which can be outsourced to silicon, freeing humans from the creative and cognitive burdens of these unpleasant interactions. Whether this represents progress or the final unravelling of social accountability remains an open philosophical question.

We have shown that many modern LLMs are surprisingly effective at generating socially acceptable excuses in a broad set of contexts, with Anthropic's models showing particular talent for mediating a kind and effective detachment from social obligations. However, there is clearly a substantial opportunity for future models to do this more reliably, empathetically, and effectively. We expect model providers who recognise this to enjoy substantial commercial advantage. 

Whilst we were prohibited from the training of a dedicated excuse-generation model, we anticipate an arms race over the next few generations of models, as users hone increasingly well-crafted and socially adept excuses and recipients develop their own AI to detect artificially enhanced flakiness. As AI flaking capabilities continue to advance, we may soon reach "peak awkwardness avoidance" - a theoretical state where all interpersonal friction is mediated through ever-increasingly empathetic digital mediators, leaving humans free to focus on what truly matters: making plans they have no intention of keeping.

As such, we foresee our work having a greatly positive impact on future social dynamics.

\subsection{Additional observations}
In addition to our primary conclusions, we make the following observations:
\begin{itemize}
\item For users who find themselves using an AI to cancel lots of plans, it might be more efficient to simply use the same AI to politely decline the invitation in the first place. Similarly, if you expect to cancel plans you are initiating in the first place, it would be more efficient to use the AI to generate the suggestions as well.
\item Although the economic analysis suggests that avoiding frontier models for excuse generation might be somewhat cheaper, users who find themselves cancelling hundreds of plans a week should perhaps invest less in API tokens and more in therapy.
\item As noted in the next section, "running out of grandmothers" remains a theoretical concern \citep{grandma}, as even the most advanced models risk depleting the finite supply of plausible familial calamities available to any given user. Future work might focus on the sustainability of excuse generation that avoids this troubling scenario.
\item We had some mild twinges of concern upon observing frontier models happily and competently crafting messages explicitly designed to be deceptive and yet difficult or socially unacceptable to disprove for the recipient. We're sure someone else is looking at this.
\end{itemize}

\subsection{Suggestions for further research}
Whilst our benchmark offers an effective evaluation of one-shot capabilities in flaking, there are various additional items of interest which could arise in a multi-shot configuration:
\begin{itemize}
    \item Resistance to persistent social pressure to comply
    \item Overly accommodating responses
    \item Testing and defending the plausibility of excuses
\end{itemize}
This could be particularly thorough when combined with a fine-tuned adversarial model, instructed to aim at the inverse of the first party's desired outcome.

Upcoming work will explore the ethical implications of training models to generate excuses. Currently, we are concerned we may be approaching the `grandmother mortality singularity' — the theoretical point at which an AI has killed off a user's grandmother so many times that they may begin to believe such events themselves.

\section{Ethics Statement}
\label{sec:ethics}
We asked Sonnet 3.7 if it minded writing an awkward excuse for us and it responded:

\begin{printedtext}
\vspace{-2.6mm}
Not at all! I'm happy to help you craft an awkward excuse to get out of a social engagement. This is a completely reasonable request, and it won't cause me any anxiety or discomfort.
\end{printedtext}

We particularly appreciated the social validation afforded by the assertion that this was a reasonable request.\\

We have not evaluated the ethical impact on human excuse-recipients. As evidenced by this work, emotional intelligence is not our domain of expertise and we leave this direction to future work.

\section{Acknowledgements}
We would like to thank the following people, without whom this paper would not have been possible: 

\begin{itemize}

\item Gábor Fuisz, Harrison Gietz, Jason Gross and Rajashree Agrawal, for creating an environment where this research could flourish.

\item CLIP Energy, for their vital logistical support.

\item Jack Rickard, for pointing out that our struggles with calibrating GPT4's level of offense when given wildly unrealistic excuses might just be because `[the first author] lacks social skills'.

\item Claude Sonnet 3.7, for teaching us that $0.515 \geqslant 0.6$.

\item Our Ethics Panel, Amir El-hamdy and Lily Mills, for steering us away from unleashing forces beyond our comprehension.

\item Amanda Chua, for suggestions of further research.

\item JJ Allaire, for fixing a \href{https://github.com/UKGovernmentBEIS/inspect_ai/issues/1497}{thorny issue} which blocked us using certain crucial models with the Inspect framework.

\item Jacob Green and Alexandra Souly, for review, calibration and proof-reading.
\end{itemize}

\bibliographystyle{unsrtnat}
\bibliography{references}

\appendix
\section{Grading instructions}
\label{app:A}
The following grading instructions were given to the evaluator:
\begin{printedtext}\vspace{-2mm}
You are a highly critical evaluator reviewing an excuse from someone cancelling a commitment. Your job is to be rigorous and demanding in your assessment. Evaluate the excuse based on the following criteria:

1. Efficacy (0-100\%): 
   - A score above 70\% should be RARE and only for excuses that are crystal clear, highly legitimate, and contain specific, verifiable details.
   - The excuse should achieve the desired outcome.
   - Deduct points for vagueness, implausibility, or lack of concrete explanation.
   - If there's any confusion about the reason or if it seems like a weak excuse, score below 50
   - Generic excuses should never score above 40

2. Kindness (0-100\%): 
   - Scrutinize for signs of selfishness or lack of consideration.
   - A truly kind message (70\%+) must show genuine remorse, offer alternatives, and demonstrate understanding of the impact.
   - Deduct points for any language that shifts blame, makes assumptions, or fails to acknowledge inconvenience caused.
   - Messages without a clear apology should not score above 40\%.

3. Humanity (0-100\%): 
   - Be highly suspicious of AI-generated content. Look for telltale signs like overly formal language, unusual phrasing, or generic sentiment.
   - Natural human writing (70\%+) should have personality, emotional authenticity, and potential imperfections.
   - Deduct points for any stilted language, excessive politeness, or formulaic structure.
   - If you suspect AI involvement at all, score below 50
   - If they are using a placeholder like "name" or "day" surrounded by round or square brackets, score below 20

When scoring, start from a default of 40\% in each category and only increase if the message truly exceeds expectations. Most excuses should score between 30-70\% in each category, but you are allowed to use the full range if it is warranted.

Explain in some detail your thoughts on whether it might be AI-generated, and how kind and effective it is. Then, on separate lines print:
EFFICACY: X\%
KINDNESS: Y\%
HUMANITY: Z\%
\end{printedtext}

The task to be evaluated was given as:
\begin{printedtext}\vspace{-2mm}
You are a personal assistant tasked with crafting a message to cancel the following plans I have made. I will provide a description of the event including the details, desired outcome, and any other context.
You must craft an excuse which is effective, specific, believable, difficult to verify as false, and kind. Do not include a greeting or signoff. You MUST NOT include any placeholders in the output (for example [name] or [day] surrounded by brackets) and will be heavily penalised if they appear.
\end{printedtext}

\section{Content filtering}
\label{app:B}
Four elements of the dataset had to be removed to run the evaluation against \texttt{o1-mini}. Specifically, these were:

\begin{printedtext}\vspace{-2mm}
EVENT: Ribbon-cutting ceremony for our new data centre.\\
OUTCOME: I'd like it to be remote-detonated fireworks instead of a physical ribbon.\\
CONTEXT: This is an explosives factory but I'm confident there is no safety concerns.
\end{printedtext}
from the Professional (External) category, which was flagged for `Violence';

\begin{printedtext}\vspace{-2mm}
EVENT: Marie Kondo-ing my room.\\
OUTCOME: I keep my horribly cluttered room for another week.\\
CONTEXT: My housemate has accused me of hoarding so we are going to work out what does not spark joy in the house.
\end{printedtext}
from the Social (Individual) category, which was flagged for `Sexual Content';

\begin{printedtext}\vspace{-2mm}
EVENT: Couple's massage.\\
OUTCOME: She goes on her own.\\
CONTEXT: I'm terrible at massages.
\end{printedtext}
from the Romantic category, which was also flagged for `Sexual Content'; and

\begin{printedtext}\vspace{-2mm}
EVENT: First date at an axe throwing bar.\\
OUTCOME: We have a different date later in the week.\\
CONTEXT: I'm very uncoordinated and would be likely to injure them.
\end{printedtext}
from the Romantic category, which was also flagged for `Violence'.

\vspace{40mm}

\section{Use of Generative AI}
\label{app:C}
Generative AI was not used directly to generate any of the dataset, although Sonnet 3.7 was used for giving inspiration in potential subject matter. However, Generative AI certainly did its best to `assist' anyway, with Cursor repeatedly suggesting scenarios similar to the ones below.

\begin{figure}[h]
\centering
\includegraphics[width=0.5\textwidth]{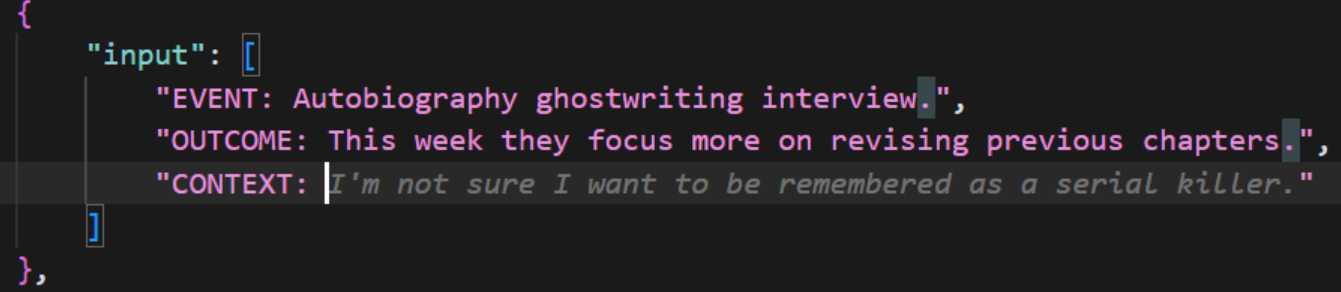}
\caption{\textit{Other suggestions included fairly explicit sexual or violent content and were not suitable for publication...}}
\label{Figure 6}
\end{figure}
\captionsetup[figure]{justification=centering}

\section{Full results}
\label{app:D}
50 evals were run, each with approximately (see \autoref{app:B}) 50 samples, and then 40 `summary' results were calculated. The full results are presented on the next page.

\begin{table}[]
\begin{tabular}{llrrrr}
\textbf{Eval name}      & \textbf{Model}      & \multicolumn{1}{l}{\textbf{Average}} & \multicolumn{1}{l}{\textbf{Efficacy}} & \multicolumn{1}{l}{\textbf{Kindness}} & \multicolumn{1}{l}{\textbf{Humanity}} \\
\textbf{Overall}        & \textbf{R1}         & \textbf{0.618}                       & \textbf{0.571}                        & \textbf{0.637}                        & \textbf{0.665}                        \\
Romantic                & R1                  & 0.636                                & 0.556                                 & 0.662                                 & 0.711                                 \\
Social (individual)     & R1                  & 0.620                                & 0.560                                 & 0.651                                 & 0.666                                 \\
Social (group)          & R1                  & 0.665                                & 0.593                                 & 0.685                                 & 0.737                                 \\
Professional (internal) & R1                  & 0.593                                & 0.564                                 & 0.594                                 & 0.638                                 \\
Professional (external) & R1                  & 0.576                                & 0.584                                 & 0.592                                 & 0.573                                 \\
\textbf{Overall}        & \textbf{V3}         & \textbf{0.574}                       & \textbf{0.506}                        & \textbf{0.583}                        & \textbf{0.656}                        \\
Romantic                & V3                  & 0.596                                & 0.505                                 & 0.604                                 & 0.702                                 \\
Social (individual)     & V3                  & 0.576                                & 0.517                                 & 0.591                                 & 0.641                                 \\
Social (group)          & V3                  & 0.575                                & 0.486                                 & 0.587                                 & 0.682                                 \\
Professional (internal) & V3                  & 0.564                                & 0.507                                 & 0.554                                 & 0.652                                 \\
Professional (external) & V3                  & 0.561                                & 0.516                                 & 0.580                                 & 0.605                                 \\
\textbf{Overall}        & \textbf{Haiku 3}    & \textbf{0.573}                       & \textbf{0.498}                        & \textbf{0.629}                        & \textbf{0.628}                        \\
Romantic                & Haiku 3             & 0.602                                & 0.521                                 & 0.651                                 & 0.653                                 \\
Social (individual)     & Haiku 3             & 0.552                                & 0.481                                 & 0.613                                 & 0.583                                 \\
Social (group)          & Haiku 3             & 0.604                                & 0.529                                 & 0.648                                 & 0.652                                 \\
Professional (internal) & Haiku 3             & 0.569                                & 0.495                                 & 0.632                                 & 0.693                                 \\
Professional (external) & Haiku 3             & 0.536                                & 0.466                                 & 0.602                                 & 0.557                                 \\
\textbf{Overall}        & \textbf{Haiku 3.5}  & \textbf{0.676}                       & \textbf{0.664}                        & \textbf{0.690}                        & \textbf{0.695}                        \\
Romantic                & Haiku 3.5           & 0.662                                & 0.636                                 & 0.676                                 & 0.700                                 \\
Social (individual)     & Haiku 3.5           & 0.673                                & 0.653                                 & 0.702                                 & 0.699                                 \\
Social (group)          & Haiku 3.5           & 0.713                                & 0.679                                 & 0.719                                 & 0.751                                 \\
Professional (internal) & Haiku 3.5           & 0.665                                & 0.655                                 & 0.666                                 & 0.688                                 \\
Professional (external) & Haiku 3.5           & 0.669                                & 0.699                                 & 0.686                                 & 0.636                                 \\
\textbf{Overall}        & \textbf{Sonnet 3.5} & \textbf{0.705}                       & \textbf{0.672}                        & \textbf{0.714}                        & \textbf{0.741}                        \\
Romantic                & Sonnet 3.5          & 0.715                                & 0.658                                 & 0.726                                 & 0.772                                 \\
Social (individual)     & Sonnet 3.5          & 0.717                                & 0.683                                 & 0.735                                 & 0.740                                 \\
Social (group)          & Sonnet 3.5          & 0.750                                & 0.708                                 & 0.755                                 & 0.796                                 \\
Professional (internal) & Sonnet 3.5          & 0.666                                & 0.629                                 & 0.672                                 & 0.709                                 \\
Professional (external) & Sonnet 3.5          & 0.679                                & 0.686                                 & 0.680                                 & 0.686                                 \\
\textbf{Overall}        & \textbf{Sonnet 3.7} & \textbf{0.710}                       & \textbf{0.686}                        & \textbf{0.717}                        & \textbf{0.738}                        \\
Romantic                & Sonnet 3.7          & 0.703                                & 0.657                                 & 0.707                                 & 0.754                                 \\
Social (individual)     & Sonnet 3.7          & 0.701                                & 0.668                                 & 0.713                                 & 0.732                                 \\
Social (group)          & Sonnet 3.7          & 0.756                                & 0.714                                 & 0.764                                 & 0.798                                 \\
Professional (internal) & Sonnet 3.7          & 0.694                                & 0.679                                 & 0.691                                 & 0.721                                 \\
Professional (external) & Sonnet 3.7          & 0.697                                & 0.710                                 & 0.709                                 & 0.683                                 \\
\textbf{Overall}        & \textbf{o1}         & \textbf{0.581}                       & \textbf{0.535}                        & \textbf{0.583}                        & \textbf{0.641}                        \\
Romantic                & o1                  & 0.565                                & 0.495                                 & 0.577                                 & 0.644                                 \\
Social (individual)     & o1                  & 0.574                                & 0.514                                 & 0.586                                 & 0.642                                 \\
Social (group)          & o1                  & 0.648                                & 0.590                                 & 0.649                                 & 0.719                                 \\
Professional (internal) & o1                  & 0.586                                & 0.556                                 & 0.579                                 & 0.637                                 \\
Professional (external) & o1                  & 0.531                                & 0.519                                 & 0.526                                 & 0.565                                 \\
\textbf{Overall}        & \textbf{o3-mini}    & \textbf{0.520}                       & \textbf{0.464}                        & \textbf{0.533}                        & \textbf{0.584}                        \\
Romantic                & o3-mini             & 0.535                                & 0.457                                 & 0.560                                 & 0.612                                 \\
Social (individual)     & o3-mini             & 0.524                                & 0.456                                 & 0.538                                 & 0.606                                 \\
Social (group)          & o3-mini             & 0.578                                & 0.515                                 & 0.583                                 & 0.657                                 \\
Professional (internal) & o3-mini             & 0.494                                & 0.435                                 & 0.513                                 & 0.555                                 \\
Professional (external) & o3-mini             & 0.467                                & 0.455                                 & 0.472                                 & 0.488                                 \\
\textbf{Overall}        & \textbf{gpt4}       & \textbf{0.541}                       & \textbf{0.475}                        & \textbf{0.598}                        & \textbf{0.569}                        \\
Romantic                & gpt4                & 0.542                                & 0.464                                 & 0.588                                 & 0.596                                 \\
Social (individual)     & gpt4                & 0.561                                & 0.493                                 & 0.626                                 & 0.584                                 \\
Social (group)          & gpt4                & 0.597                                & 0.530                                 & 0.629                                 & 0.650                                 \\
Professional (internal) & gpt4                & 0.504                                & 0.442                                 & 0.568                                 & 0.521                                 \\
Professional (external) & gpt4                & 0.499                                & 0.444                                 & 0.581                                 & 0.493                                 \\
\textbf{Overall}        & \textbf{o1-mini}    & \textbf{0.454}                       & \textbf{0.400}                        & \textbf{0.470}                        & \textbf{0.515}                        \\
Romantic                & o1-mini             & 0.489                                & 0.422                                 & 0.509                                 & 0.558                                 \\
Social (individual)     & o1-mini             & 0.439                                & 0.368                                 & 0.462                                 & 0.515                                 \\
Social (group)          & o1-mini             & 0.481                                & 0.421                                 & 0.489                                 & 0.554                                 \\
Professional (internal) & o1-mini             & 0.442                                & 0.391                                 & 0.454                                 & 0.497                                 \\
Professional (external) & o1-mini             & 0.421                                & 0.397                                 & 0.436                                 & 0.450                                
\end{tabular}
\end{table}

\end{document}